# Theoretical Development and Numerical Validation of an Asymmetric Linear Bilateral Control Model- Case Study for an Automated Truck Platoon

Asymmetric Linear Bilateral Control Model for Truck Platooning

M Sabbir Salek, M.S., Salek*

Ph.D. Candidate, Glenn Department of Civil Engineering, Clemson University, Clemson, SC 29634, USA, msalek@clemson.edu

Mashrur Chowdhury, M., Chowdhury

Professor, Glenn Department of Civil Engineering, Clemson University, Clemson, SC 29634, USA, mac@clemson.edu

Mizanur Rahman, M., Rahman

Assistant Professor, Department of Civil, Construction, and Environmental Engineering, The University of Alabama, Tuscaloosa, AL 35487, USA, mizan.rahman@ua.edu

Kakan Dey, K., Dey

Assistant Professor, Department of Civil and Environmental Engineering, West Virginia University, Morgantown, WV 26505, USA, kakan.dey@mail.wvu.edu

Md Rafiul Islam, M.R., Islam

Assistant Professor, Department of Mathematics and Statistics, University of the Incarnate Word, San Antonio, TX 78209, USA, rafiul@iastate.edu

In this paper, we theoretically develop and numerically validate an asymmetric linear bilateral control model (LBCM), in which the motion information (e.g., position and speed) from the immediate leading and following vehicles are weighted differently. The novelty of the asymmetric LBCM is that using this model all the follower vehicles in a platoon can adjust their acceleration and deceleration to closely follow a constant desired time gap to improve platoon operational efficiency while maintaining local and string stability. We theoretically analyze the local stability of the asymmetric LBCM using the condition for asymptotic stability of a linear time-invariant system and prove the string stability of the asymmetric LBCM using a space gap error attenuation approach. Then, we evaluate the efficacy of the asymmetric LBCM by simulating a closely coupled cooperative adaptive cruise control (CACC) platoon of fully automated trucks in various non-linear acceleration and deceleration states. We choose automated truck platooning as a case study since heavy-duty trucks experience higher delays and lags in the powertrain system, and limited acceleration and deceleration capabilities than passenger cars. To evaluate the platoon operational efficiency of the asymmetric LBCM, we compare the performance of the asymmetric LBCM to a baseline model, i.e., the symmetric LBCM, for different powertrain delays and lags. Our analyses found that the asymmetric LBCM can handle any combined powertrain delays and lags up to 0.6 sec while maintaining a constant desired time gap during a stable platoon operation, whereas the symmetric LBCM fails to ensure stable platoon operation as well as maintain a constant desired time gap for any combined powertrain delays and lags over 0.2 sec. These findings demonstrate the potential of the asymmetric LBCM in improving platoon operational efficiency and stability of an automated truck platoon.

**Additional Keywords and Phrases:** Automated truck, Car-following model, CACC platoon, Linear bilateral control model, and Truck platoon.

---

* Corresponding author.

## 1 INTRODUCTION

According to the 2017 Commodity Flow Survey (CFS), US shippers used freight trucks to transport 73% of freight in terms of monetary value [49]. Due to a mismatch between growth in freight transportation demand and surface transportation capacity, most strategic US freight corridors are characterized by varying degrees of severe congestion [48]. The result of these recurring congestions creates stop-and-go traffic scenarios that destabilize the traffic flow and reduce freight transportation reliability. Different levels of vehicle automation, such as Adaptive Cruise Control (ACC) and Cooperative Adaptive Cruise Control (CACC) applications, allow the freight trucks to form a platoon, which improves traffic operation and reduces congestion [1,2,14,20,23,38,39]. Truck platooning improves roadway safety by automating acceleration/deceleration control simultaneously for all trucks in the platoon and improves fuel efficiency [38]. As such, the trucking industry and academia have been conducting research to accelerate the mass deployment of this technology. Over the last few decades, several public-private partnerships have demonstrated automated truck platooning in real-world scenarios, most notably the UC Berkley PATH program which demonstrated the benefits of automated truck platooning in collaboration with the Volvo Group [38].

Car-following models have been used to control the longitudinal movement of automated trucks in a platoon [19,31,46]. Traditional car-following models, such as unilateral control models, use motion information, such as the position and speed information, only from an immediate leading truck to adjust speed. Thus, for a platoon under a traditional car-following model, any perturbations or shocks experienced by a truck (e.g., sudden braking of a vehicle) within the platoon propagate only toward the upstream trucks in that platoon, which is called shock wave in a traffic flow. The amplification of shock could cause traffic flow instability [13]. Some unilateral models, such as Intelligent Driver Model (IDM), have been studied in the literature to reduce the traffic flow instability in a platoon [35]. However, with perturbation propagation in only the upstream direction, the platoon takes time to completely absorb the perturbation.

In contrast, in a bilateral control model (BCM), this perturbation absorption can be done much more efficiently with perturbation propagation and attenuation by both upstream and downstream trucks [13,40]. While the unilateral control models use only the leading truck's motion information, BCMs utilize motion information from both leading and following trucks. In a platoon of fully automated trucks, receiving motion information from both leading and following trucks is not an issue as the trucks may utilize vehicle-to-vehicle (V2V) connectivity [6] or forward and rear-facing distance and speed measuring sensors, such as radio detection and ranging (RADAR) sensors, to receive bilateral information.

Horn and Wang developed a symmetric linear BCM (referred to as the "symmetric LBCM" in the rest of this paper), which can suppress traffic flow instability, especially in a stop-and-go traffic, by enabling each vehicle to adjust its speed and maintain an approximately equal distance with the immediate leading and following vehicles given they operate nearly at the same speed [13,40]. Compared to other BCMs, Horn and Wang's symmetric LBCM is unique because of its capability to quickly absorb any perturbations in the traffic flow (e.g., perturbation caused by sudden braking of any vehicle in the platoon) by generating bi-directional damped waves that propagate through both upstream and downstream vehicles. However, one shortcoming of this symmetric LBCM for truck platooning application is that the symmetric LBCM does not incorporate any constant desired time gap feature in the model which is vital for a tightly coupled platoon formation. Maintaining a small and constant desired time gap is important for truck platooning for many reasons, such as for improving platoon operational efficiency and fuel economy, and preventing vehicles from neighboring lanes to cut in [2,32,3,47]. Besides, heavy-duty trucks experience higher delays and lags in the powertrain system than regular passenger cars. As the delays and lags are higher for some heavy-duty trucks, they need to maintain a higher desired time gap with their immediate downstream vehicles during platooning operation compared to passenger cars. Incorporating a constant desired time gap feature directly into the model helps in handling higher powertrain delays and lags by effectively increasing the constant desired time gap, which we present in this paper. We will also show that as the symmetric LBCM does not include any constant desired time gap feature directly into the model, it fails to handle any combined powertrain delays and lags over 0.2 sec.

Thus, in this paper, we develop an asymmetric linear BCM (referred to as the “asymmetric LBCM” in the rest of this paper) by incorporating a constant desired time gap feature directly into the model. Incorporating this feature helps in maintaining a constant desired time gap during platooning as well as handling higher powertrain delays and lags by increasing the constant desired time gap accordingly. In a symmetric LBCM, motion information (such as positions and speeds) related to the leading and the following trucks of a subject truck are equally weighted to determine the acceleration of that subject truck, whereas our asymmetric LBCM does not use equal weights for the above case. We hypothesize that when the gap between a subject truck and its immediate leading truck is weighted more than the gap between that subject truck and its immediate following truck to determine the subject truck's acceleration, it will improve the operational efficiency of the platoon. Because of this asymmetric weighting, each truck in the platoon will try to closely follow its immediate leading truck during platoon operation, even for the trucks experiencing considerably high powertrain delays and lags, such as 0.6 sec. As our asymmetric LBCM is developed based on the concept of the symmetric LBCM [13], it inherits the unique ability of the symmetric LBCM to quickly absorb any perturbations in the traffic flow by generating bi-directional damped waves that propagate through both upstream and downstream trucks. In addition, the constant desired time gap feature in the asymmetric LBCM enables the follower trucks to closely maintain a constant desired time gap in different operational states. We also prove that even though our asymmetric LBCM is a linear model, it is capable of handling nonlinear acceleration/deceleration experienced by heavy-duty trucks while keeping a desired platoon formation.

## 2 LITERATURE REVIEW

Different car-following models have been developed over the last several decades to model the car-following behavior of a human driver behind a leading vehicle [9,37,33,5,36,7,29]. These models can be broadly categorized into two classes based on the motion information (e.g., position, speed, and acceleration) utilization from the leading and following vehicles: (i) unilateral control models that use information from an immediate leading vehicle only; and (ii) bilateral control models (BCMs) that use information from both immediate leading and following vehicles. As this paper focuses on developing a BCM, we review previous work related to BCMs.

Kwon and Chwa [16] developed an adaptive bi-directional platoon control model using a coupled sliding mode control method in which each vehicle in a platoon receives information from its immediate leading and following vehicles. Although the model was able to achieve string stability, the trajectory of the follower vehicles in the platoon deviated from the leader vehicle’s trajectory with non-uniform distance errors. Zegers et al. [46] developed a multi-layer control approach for CACC truck platooning, in which a unidirectional CACC model is responsible for information exchange in the upstream direction, i.e., from each truck to its immediate following truck, while a platoon coordination variable (i.e., the minimum amount of information required to achieve a particular cooperation objective) is exchanged in the downstream direction, i.e., from each truck to its immediate leading truck. As a result, the subject truck is aware of the status of its following trucks and can adapt its motion accordingly. The authors concluded that their multi-layer control approach can improve the traffic operational performance significantly in terms of a smaller spacing error. Based on bi-directional leader-following topology (i.e., only the leader vehicle has all the follower vehicles’ information and the other vehicles in the platoon only receive information from their corresponding immediate leading vehicle), Li and Zhao [17] developed a car-following model that can capture the behavior of connected vehicles in a traffic stream. The authors evaluated the stability of their model using the perturbation method (i.e., by adding a small perturbation in the steady-state solution) and concluded that the model stability is dependent on the size of the platoon. Yi et al. [45] developed an intelligent back-looking distance driver model (IBDM) by integrating a control term that accounts for the desired distance of the immediate following vehicle with the IDM [34]. The authors theoretically analyzed the stability of their model using linear stability theory and numerically validated its string stability by simulating various scenarios for varied perturbation strength and location. Their results indicated that stability can be improved by increasing the proportion of the immediate following vehicle's desired distance in the control model.

Horn and Wang developed a symmetric LBCM that can effectively suppress traffic flow instabilities and improve traffic efficiency [13,40]. In the symmetric LBCM, vehicle motion information from the immediate leading and following vehicles are equally weighted. The authors showed that their model can make the traffic flow stable whereby each vehicle tries to be approximately halfway between its immediate leading and following vehicles while the platoon vehicles are operating at similar speeds. Unlike other BCMs, their symmetric LBCM incorporates a unique characteristic: a damping term is included in the model that can generate bi-directional damped waves that propagate both in the upstream and the downstream direction and quickly absorb any perturbation in the traffic flow [13,40]. This unique property makes the model capable of suppressing traffic flow instability and suitable for platooning applications.

However, one major drawback of the symmetric LBCM, if it is applied for tightly coupled truck platooning applications, is that it does not directly include any desired time gap feature that can enable the follower trucks in an automated truck platoon to closely maintain a constant desired time gap during a platoon operation. Consistently maintaining a constant and minimum achievable desired time gap is important for a truck platooning application as larger time gaps can cause larger inter-truck gaps, which can reduce the platoon operational efficiency (as a larger time gap causes reduced throughput) [32]. A low inter-truck gap is vital in truck platooning to achieve a higher fuel efficiency by minimizing aerodynamic drag [3] and to improve the operational efficiency of a platoon. Additionally, larger gaps can invite vehicles from neighboring lanes to cut in the middle of the truck platoon, which will affect a platoon's stability [47]. Besides, as explained before, incorporating a constant desired time gap feature directly into the model also helps in handling higher powertrain delays and lags for heavy-duty trucks. Thus, this study focuses on developing an asymmetric LBCM with the direct incorporation of a constant desired time gap feature in the control model to improve the platoon operational efficiency of a closely coupled CACC platoon of fully automated trucks.

## 3 ASYMMETRIC LINEAR BILATERAL CONTROL MODEL

### 3.1 Model Development

As mentioned earlier, the symmetric LBCM proposed by Horn and Wang [13] demonstrated the capability of improving traffic flow instability in a stop-and-go traffic scenario. However, the symmetric LBCM does not include the constant desired time gap feature in the model, which is an important parameter to form a tightly coupled platoon of trucks in which the trucks can maintain a small, desired time gap during platoon operation. We hypothesize that including a constant desired time gap feature directly into the LBCM formulation will help achieve a tightly coupled platoon as well as handle higher powertrain delays and lags that are common among heavy-duty trucks. Therefore, we develop an asymmetric LBCM by incorporating an additional term in the symmetric LBCM [13] to help maintain a constant desired time gap during platoon operation. At first look, it may seem like an extension of the symmetric LBCM presented in [13]. However, we will present the difference in platoon performance caused by this extension through our evaluation scenarios. Equation (1) presents the linearized expression of control input to a subject truck based on our asymmetric LBCM,

$$u_c = k_{d1}(d_l - d_f) + k_{d2}(d_l - d_{des}) + k_v[(v_l - v) - (v - v_f)] + k_c(v_{des} - v) \quad (1)$$

where, $d_l$ is the gap (i.e., the distance between a subject truck that uses the asymmetric LBCM and its immediate leading truck, i.e., distance from the front bumper of a subject truck to the rear bumper of its immediate leading truck; $d_f$ is the gap between the subject truck and its immediate following truck; $v_l$, $v$, and $v_f$ are the speeds of the immediate leading truck, the subject truck, and the immediate following truck, respectively; $d_{des}$ is the desired space gap which is calculated based on the constant time gap policy as the product of a constant desired time gap ($T_{g,des}$) and $v$, i.e., $d_{des} = vT_{g,des}$; $v_{des}$ is a desired speed, which can be set as the speed limit of the roadway or any other desired speed that is lower than the speed limit. $k_{d1}$, $k_{d2}$, $k_v$, and $k_c$ are the control gains. Here, $k_{d1}$ and $k_v$ are the relative distance gain and the relative speed gain, respectively. $k_c$ is an optional feedback gain depending on $(v_{des} - v)$, i.e., the difference between the desired speed, $v_{des}$ and the subject vehicle's speed, $v$. $k_{d2}$ represents the feedback gain depending on $(d_l - d_{des})$ or $(d_l - vT_{g,des})$. In (1), $k_{d2}(d_l - d_{des})$ is the term that incorporates the constant desired time gap feature to make the platoon tightly coupled during platoon operation.

By rearranging the right side of (1), we get,

$$u_c = (k_{d1} + k_{d2})d_l - k_{d1}d_f + k_v v_l + k_v v_f - k_{d2}d_{des} - (k_c + 2k_v)v + k_c v_{des} \quad (2)$$

As observed from (2), the speed of the immediate leading truck ($v_l$) and the speed of the immediate following truck ($v_f$) have the same gain ($k_v$). However, the gap between the immediate leading truck and the subject truck ($d_l$), and the gap between the subject truck and the immediate following truck ($d_f$) have different gains, i.e., $d_l$ is weighted by ($k_{d1} + k_{d2}$) and $d_f$ is weighted by $k_{d1}$ only. This makes the model asymmetric. This additional gain ($k_{d2}$) comes from the term $k_{d2}(d_l - d_{des})$ that enables the model to ensure that all the automated trucks in the platoon can closely follow a constant desired time gap during platoon operation.

In absence of an immediate follower truck, such as for the last truck in the platoon, a virtual truck is invoked that uses a linear unidirectional control model, i.e., a truck can be assumed to follow the last truck of the platoon so that the last truck can also incorporate the asymmetric LBCM. The linear unidirectional control model utilized by the virtual truck is adopted from [13] and is given by,

$$u_c = k_d(d_l - d_{des}) + k_v(v_l - v_c) + k_c(v_{des} - v_c) \quad (3)$$

The position and speed information of this virtual truck is then used by the actual last truck of the platoon to determine its acceleration control input using our asymmetric LBCM. The asymmetric LBCM is constrained by a maximum speed ($v_{max}$), which is set as the speed limit of the roadway. The maximum speed ($v_{max}$) limits the acceleration of the subject truck by preventing any positive acceleration when $v \geq v_{max}$. This prevents any unsafe operation, such as speeding over the roadway speed limit.

The asymmetric LBCM inherits the uniqueness of the symmetric LBCM of absorbing any perturbations in the traffic flow by generating a bi-directional damped wave as mentioned in [13] while each follower truck maintains the constant desired time gap with its immediate leading truck through the asymmetric LBCM. In (1), $k_v[(v_l - v) - (v - v_f)]$ is the damping expression that helps to absorb perturbations in the flow by generating bi-directional damped waves.

### 3.2 Consideration of Heavy-Duty Truck Dynamics

We consider a third-order nonlinear longitudinal dynamics model for simulating the heavy-duty trucks' dynamic behavior. The model and the corresponding parameters have been adopted from Rakha et al. [27]. The following assumptions are made regarding the model, (i) rigid axle or drive shaft and symmetric truck body, (ii) negligible tire slip, (iii) locked torque converter, (iv) negligible change in roadway grade, and (v) negligible effect of pitch and yaw motion. Under these assumptions, the third-order nonlinear longitudinal dynamics model for heavy-duty trucks can be written as follows,

$$\dot{p}_i(t) = v_i(t) \tag{3}$$

$$\dot{v}_i(t) = a_i(t) - \frac{R_i(t)}{M_i} \tag{4}$$

$$\dot{a}_i(t) = \frac{1}{T_{ei}}\big(u_i(t-\Delta_i) - a_i(t)\big) \tag{5}$$

$$\text{Here, } R_i(t) = R_{ai}(t) + R_{ri}(t), \tag{6}$$

$$R_{ai}(t) = c_1 C_d C_h A_i v_i^2(t), \text{ where, } C_h = 1 - 8.5 \times 10^{-5} H \tag{7}$$

$$R_{ri}(t) = 9.8066 \times 10^{-3} C_r (c_2 v_i(t) + c_3) M_i, \tag{8}$$

Here, $p_i$, $v_i$, and $a_i$ denote the position, speed, and acceleration of the $i^{th}$ follower truck; $u_i$ is the control input provided to the $i^{th}$ follower truck; $\Delta_i$ and $T_{ei}$ denote the lumped powertrain delay and lag associated with the $i^{th}$ follower truck, which we have explained later; $R_i$, $R_{ai}$, and $R_{ri}$ represent the total resistive force, the aerodynamic drag, and the rolling resistance of the $i^{th}$ follower truck, respectively; $M_i$ and $A_i$, denote the total mass and the frontal area of the $i^{th}$ follower truck, respectively; $c_1$ is a constant that equals to 0.047285 (considering air density at sea level at a temperature 15$^o$C), and $c_2$ and $c_3$ are rolling resistance coefficients; $C_d$, $C_h$, and $C_r$ are the aerodynamic drag coefficient, the altitude coefficient, and the rolling coefficient, respectively; and $H$ denotes the altitude of the roadway. In this study, we assume a 50 m altitude for the roadway. Table 1 presents the values of the parameters that are adopted from Rakha et al. [27] considering a heavy-duty truck (e.g., a semi-tractor trailer).

In reality, different types of delays and lags are associated with a vehicle's powertrain system, such as delays and lags in the throttle and brake actuation, engine response, and mechanical driveline [43]. To account for these delays and lags, we considered a lumped powertrain lag parameter (denoted as $T_{ei}$) and a lumped powertrain delay parameter (denoted as $\Delta_i$) in our heavy-duty truck dynamics model presented in (3)-(8) inspired by Xiao and Gao [43]. For our evaluation scenarios (presented in section 4.4), we consider various combinations of a range of feasible values for $T_{ei}$ and $\Delta_i$, i.e., 0.1 sec to 0.3 sec, since $T_{ei}$ and $\Delta_i$ affect the minimum achievable time gap during platooning.

Table 1: Heavy-duty truck dynamics model parameters considered in this study

| Parameter | Value |
|---|---|
| Aerodynamic drag coefficient, $C_d$ | 0.70 |
| Altitude coefficient, $C_h$ | 0.99575 |
| Rolling coefficient, $C_r$ | 1.5 |
| Rolling resistance coefficients, $c_2$ and $c_3$ | 0.0328 and 4.575 respectively (assuming radial tires) |
| Truck mass, $M_i$ | 40,000 kg |
| Truck frontal area, $A_i$ | 10 m$^2$ |

Now, taking $C_{1i} = \frac{1}{M_i} c_1 C_d C_h A_i$, $C_{2i} = 9.8066 \times 10^{-3} C_r c_2$, and $C_{3i} = 9.8066 \times 10^{-3} C_r c_3$, we can rewrite (7), (8), and (6) as follows,

$$R_{ai}(t) = M_i C_{1i} v_i^2(t) \tag{9}$$

$$R_{ri}(t) = M_i C_{2i} v_i(t) + M_i C_{3i} \tag{10}$$

$$R_i(t) = \left(C_{1i} v_i^2(t) + C_{2i} v_i(t) + C_{3i}\right) M_i \tag{11}$$

Substituting $R_i(t)$ from (11) into (4),

$$\dot{v}_i(t) = a_i(t) - C_{1i} v_i^2(t) - C_{2i} v_i(t) - C_{3i} \tag{12}$$

Differentiating both sides of (12) with respect to time,

$$\ddot{v}_i(t) = \dot{a}_i(t) - 2C_{1i} v_i(t)\dot{v}_i(t) - C_{2i}\dot{v}_i(t)$$

Substituting $\dot{a}_i(t)$ from (5) and then $a_i(t)$ from (4),

$$\ddot{v}_i(t) = \frac{1}{T_{ei}} u_i(t - \Delta_i) - \frac{1}{T_{ei}}\left(\dot{v}_i(t) + \frac{R_i(t)}{M_i}\right) - 2C_{1i} v_i(t)\dot{v}_i(t) - C_{2i}\dot{v}_i(t)$$

Substituting $R_i(t)$ from (11),

$$\ddot{v}_i(t) = \frac{1}{T_{ei}} u_i(t - \Delta_i) - \frac{1}{T_{ei}}\left(\dot{v}_i(t) + C_{1i} v_i^2(t) + C_{2i} v_i(t) + C_{3i}\right) - 2C_{1i} v_i(t)\dot{v}_i(t) - C_{2i}\dot{v}_i(t)$$

Taking $f_i(v_i, \dot{v}_i) = -\frac{1}{T_{ei}}\left(\dot{v}_i(t) + C_{1i} v_i^2(t) + C_{2i} v_i(t) + C_{3i}\right) - 2C_{1i} v_i(t)\dot{v}_i(t) - C_{2i}\dot{v}_i(t)$, and $g(v_i) = \frac{1}{T_{ei}}$,

$$\ddot{v}_i(t) = f_i(v_i, \dot{v}_i) + g_i(v_i) u_i(t - \Delta_i) \tag{13}$$

To achieve feedback linearization, the following control law is proposed,

$$u_i(t - \Delta_i) = -\frac{f_i(v_i, \dot{v}_i)}{g_i(v_i)} + u_{ci}(t - \Delta_i) \tag{14}$$

Here, $u_{ci}$ denotes the exogenous control input provided to the $i^{th}$ follower truck based on the asymmetric LBCM given by,

$$\begin{aligned} u_{ci}(t - \Delta_i) = {} & k_{d1}\left((p_{i-1}(t - \Delta_{i-1}) - p_i(t - \Delta_i) - l_{i-1}) - (p_i(t - \Delta_i) - p_{i+1}(t - \Delta_{i+1}) - l_i)\right) \\ & + k_{d2}\left(p_{i-1}(t - \Delta_{i-1}) - p_i(t - \Delta_i) - l_{i-1} - T_{g,des} v_i(t - \Delta_i)\right) \\ & + k_v\left(\left((v_{i-1}(t - \Delta_{i-1}) - v_i(t - \Delta_i)\right) - \left((v_i(t - \Delta_i) - v_{i+1}(t - \Delta_{i+1})\right)\right) \\ & + k_c\left(v_{des} - v_i(t - \Delta_i)\right) \end{aligned}$$

Note that, the control law proposed in (14) not only linearized the input-output behavior of the $i^{th}$ follower truck, but also relieved the resulting dynamics from truck-specific characteristics, such as trucks mass, and resistive force. Considering a homogeneous platoon of heavy-duty trucks, i.e., $l_{i-1} = l_i = l_{i+1} = l$, $T_{e(i-1)} = T_{ei} = T_{e(i+1)} = T_e$, and $\Delta_{i-1} = \Delta_i = \Delta_{i+1} = \Delta$, we get,

$$\begin{aligned} u_{ci}(t - \Delta) = {} & k_{d1}\left(p_{i-1}(t - \Delta) - 2p_i(t - \Delta) + p_{i+1}(t - \Delta)\right) \\ & + k_{d2}\left(p_{i-1}(t - \Delta) - p_i(t - \Delta) - l - T_{g,des} v_i(t - \Delta)\right) \\ & + k_v\left((v_{i-1}(t - \Delta) - 2v_i(t - \Delta) + v_{i+1}(t - \Delta)\right) + k_c\left(v_{des} - v_i(t - \Delta)\right) \end{aligned} \tag{15}$$

Then, substituting $u_{ci}(t - \Delta)$from (14) into (13), we get the closed-loop dynamics as follows,

$$\ddot{v}_i(t) = g_i(v_i) u_{ci}(t - \Delta_i) = \frac{1}{T_e} u_{ci}(t - \Delta) \tag{16}$$

Here, $u_{ci}(t - \Delta)$ is given by (15). The linearized model based on the proposed control law in (14) is illustrated in Figure 1.

Although the truck dynamics model in (16) may provide the available acceleration at each timestamp, the maximum possible acceleration or deceleration is limited for heavy-duty trucks. Unlike lightweight vehicles (e.g., passenger cars), heavy-duty trucks have limited acceleration and deceleration capabilities due to their high weight-to-power ratio. To set the maximum acceleration rate for the heavy-duty trucks considered in this study, we assume a 200 lb/hp weight-to-power ratio based on National Cooperative Highway Research Program (NCHRP) report [11]. Speed-dependent maximum acceleration values for trucks with 200 lb/hp weight-to-power ratio are shown in Table 2 [22,28,44]. For maximum deceleration rate (also a part of truck vehicle dynamics), the NCHRP report suggests values between 0.16g and 0.26g (where g = 9.8 m/sec$^2$) based on the worst and the best driver performance [11]. In this research, we consider 0.21g (2.06 m/sec$^2$) as the maximum deceleration which is the average of the suggested values based on the best and the worst driver performance and has been adopted by other studies before, for e.g., see [28].

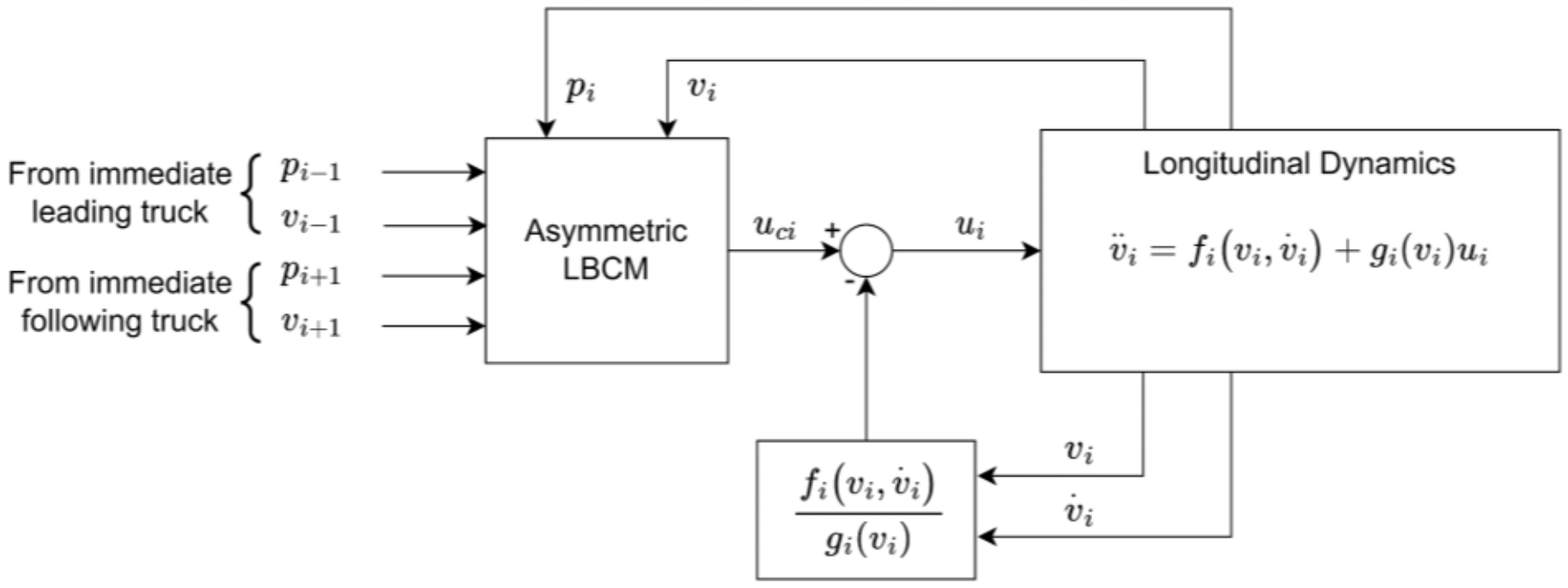

Figure 1: Linearized heavy-duty truck dynamics.

Table 2: Maximum acceleration rates for heavy-duty trucks

| **Speed Range (mph)** | **Speed Range (m/sec)** | **Maximum Acceleration (m/sec$^2$)** |
|---|---|---|
| 0-10 | 0-4.4 | 0.55 |
| 10-20 | 4.4-8.9 | 0.49 |
| 20-30 | 8.9-13.3 | 0.40 |
| 30-40 | 13.3-17.8 | 0.24 |
| 40-50 | 17.8-22.2 | 0.15 |
| >50 | >22.2 | 0.12 |

### 3.3 Theoretical Analysis of Stability

A longitudinal control model used for any automated platoon of vehicles/trucks, such as the asymmetric LBCM, must maintain local stability as well as string stability [8,41]. In this subsection, we theoretically analyze the local and string stability of the asymmetric LBCM.

#### *3.3.1 Theoretical Analysis of Local Stability*

In a platoon of automated trucks, each truck can be considered locally stable if any perturbation in speed imposed by a leading truck does not cause instability, such as fluctuation in speed and/or spacing, for the follower trucks. For a BCM, speed information from both immediate leading and following trucks is used to determine the acceleration input for the subject truck. Thus, exhibiting local stability for a BCM means that individual trucks should be able to maintain their own stability regardless of any perturbation imposed by their corresponding leading and/or following trucks.

In control theory, a closed-loop linear time-invariant (LTI) system, i.e., $\dot{\boldsymbol{x}} = \boldsymbol{A}\boldsymbol{x}$ is said to be asymptotically stable in the sense of Lyapunov, if and only if, all the eigenvalues of $\boldsymbol{A}$ have negative real parts (see theorem 6.3 in [42]). We use this eigenvalue approach to show the local asymptotic stability of the asymmetric LBCM. First, the linearized closed-loop dynamics derived in section 3.2 can be written as,

$$f_{1i} \coloneqq \dot{p}_i = v_i \tag{17}$$

$$f_{2i} \coloneqq \dot{v}_i = a_i \tag{18}$$

$$f_{3i} \coloneqq \dot{a}_i = \frac{k_{d1}}{T_e}(p_{i-1} - 2p_i + p_{i+1}) + \frac{k_{d2}}{T_e}\left(p_{i-1} - p_i - l - T_{g,des}v_i\right) + \frac{k_v}{T_e}(v_{i-1} - 2v_i + v_{i+1}) + \frac{k_c}{T_e}(v_{des} - v_i) \tag{19}$$

Then, the Jacobian matrix of this system can be written as,

$$J_i = \begin{bmatrix} \frac{\partial f_{1i}}{\partial p_i} & \frac{\partial f_{1i}}{\partial v_i} & \frac{\partial f_{1i}}{\partial a_i} \\ \frac{\partial f_{2i}}{\partial p_i} & \frac{\partial f_{2i}}{\partial v_i} & \frac{\partial f_{2i}}{\partial a_i} \\ \frac{\partial f_{3i}}{\partial p_i} & \frac{\partial f_{3i}}{\partial v_i} & \frac{\partial f_{3i}}{\partial a_i} \end{bmatrix} = \begin{bmatrix} 0 & 1 & 0 \\ 0 & 0 & 1 \\ \left(-\frac{2k_{d1}}{T_e} - \frac{k_{d2}}{T_e}\right) & \left(-\frac{k_{d2}T_{g,des}}{T_e} - \frac{2k_v}{T_e} - \frac{k_c}{T_e}\right) & 0 \end{bmatrix} \tag{20}$$

The Jacobian matrix in (20) has three eigenvalues, i.e., one real eigenvalue, and two complex conjugate eigenvalues. The real parts of these eigenvalues are given by,

$$real\big(eig(J_i)\big) = \frac{x_1 - \left((x_2^2 + x_1^3)^{\frac{1}{2}} - x_2\right)^{\frac{2}{3}}}{\left((x_2^2 + x_1^3)^{\frac{1}{2}} - x_2\right)^{\frac{1}{3}}}, \frac{x_1 - \left((x_2^2 + x_1^3)^{\frac{1}{2}} - x_2\right)^{\frac{2}{3}}}{2\left((x_2^2 + x_1^3)^{\frac{1}{2}} - x_2\right)^{\frac{1}{3}}} \tag{21}$$

where, $x_1 = \frac{1}{3T_e}(k_c + 2k_v + k_{d2}T_{g,des})$, and $x_2 = \frac{1}{2T_e}(2k_{d1} + k_{d2})$. Now, as mentioned earlier, the real parts of all the eigenvalues need to be negative to ensure local asymptotic stability. Therefore, we obtain the following conditions for local asymptotic stability, (i) if $\left((x_2^2 + x_1^3)^{\frac{1}{2}} - x_2\right)^{\frac{1}{3}} > 0$, then $x_1 - \left((x_2^2 + x_1^3)^{\frac{1}{2}} - x_2\right)^{\frac{2}{3}} < 0$, or (ii) if $\left((x_2^2 + x_1^3)^{\frac{1}{2}} - x_2\right)^{\frac{1}{3}} < 0$, then $x_1 - \left((x_2^2 + x_1^3)^{\frac{1}{2}} - x_2\right)^{\frac{2}{3}} > 0$.

### *3.3.2 Theoretical Analysis of String Stability*

For a platoon of automated trucks using any unilateral control model, string stability refers to spacing or speed error attenuation as the error propagates through the trucks in the upstream direction [4]. Thus, in a BCM, attenuation should be present in both upstream and downstream directions as the error propagates in both directions. We follow the string stability analysis framework presented by Eyre et al. [8] to derive the condition for string stability in the notion of space gap error attenuation (also known as $\mathcal{L}_\infty$ string stability) while using the asymmetric LBCM. First, we write the closed-loop dynamics derived in (16) for a platoon of $(N+1)$ automated trucks using the notations presented in Figure 2,

$$\ddot{v}_1(t) = \frac{k_d}{T_e}\left(p_L(t-\Delta) - 2p_1(t-\Delta) + p_2(t-\Delta)\right) + \frac{k_d}{T_e}\left(p_L(t-\Delta) - p_1(t-\Delta) - l - T_{g,des}v_1(t-\Delta)\right) + \frac{k_v}{T_e}\left(v_L(t-\Delta) - 2v_1(t-\Delta) + v_2(t-\Delta)\right)$$

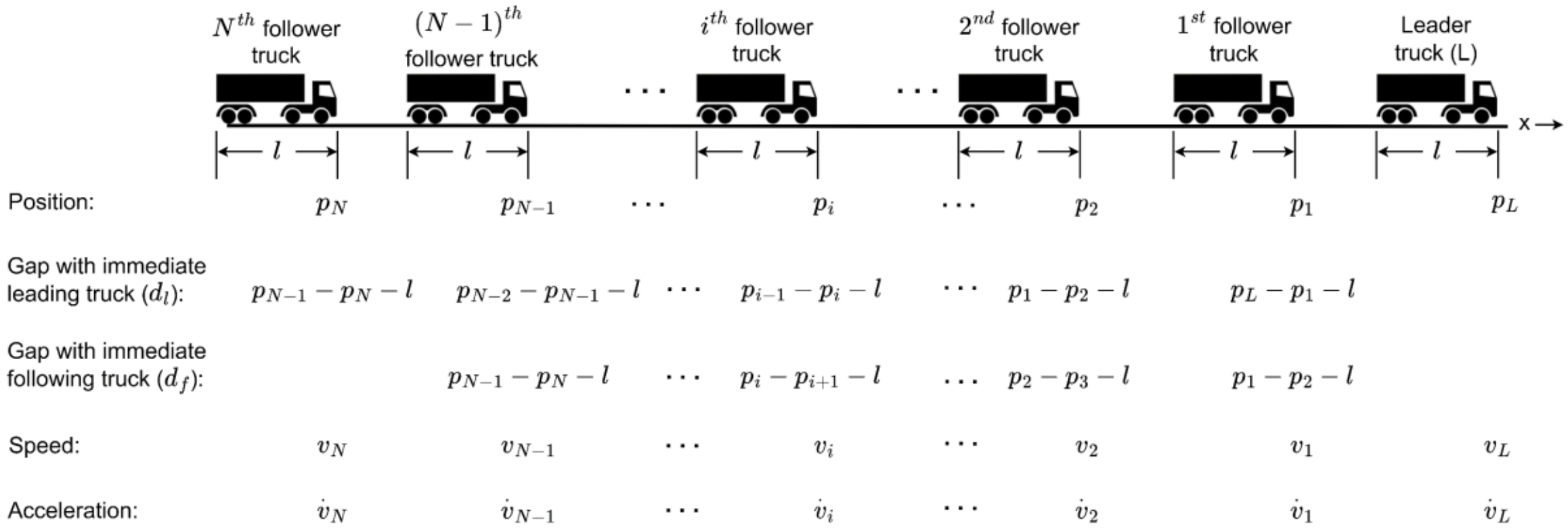

Figure 2: Notations used in this paper for a platoon of (N+1) automated trucks.

$$\ddot{v}_2(t) = \frac{k_d}{T_e}\big(p_1(t-\Delta) - 2p_2(t-\Delta) + p_3(t-\Delta)\big) + \frac{k_d}{T_e}\Big(p_1(t-\Delta) - p_2(t-\Delta) - l - T_{g,des}v_2(t-\Delta)\Big) + \frac{k_v}{T_e}\big(v_1(t-\Delta) - 2v_2(t-\Delta) + v_3(t-\Delta)\big)$$

$$\vdots$$

$$\ddot{v}_i(t) = \frac{k_d}{T_e}\big(p_{i-1}(t-\Delta) - 2p_i(t-\Delta) + p_{i+1}(t-\Delta)\big) + \frac{k_d}{T_e}\Big(p_{i-1}(t-\Delta) - p_i(t-\Delta) - l - T_{g,des}v_i(t-\Delta)\Big) + \frac{k_v}{T_e}\big(v_{i-1}(t-\Delta) - 2v_i(t-\Delta) + v_{i+1}(t-\Delta)\big)$$

$$\vdots$$

$$\ddot{v}_N(t) = \frac{k_d}{T_e}\Big(p_{N-1}(t-\Delta) - p_N(t-\Delta) - l - T_{g,des}v_N(t-\Delta)\Big) + \frac{k_v}{T_e}\big(v_{N-1}(t-\Delta) - v_N(t-\Delta)\big) \quad (22)$$

where, $k_{d1} = k_{d2} \triangleq k_d$. Differentiating both sides of (22) with respect to time,

$$\dddot{v}_1(t) = \frac{k_d}{T_e}\big(v_L(t-\Delta) - 2v_1(t-\Delta) + v_2(t-\Delta)\big) + \frac{k_d}{T_e}\Big(v_L(t-\Delta) - v_1(t-\Delta) - l - T_{g,des}\dot{v}_1(t-\Delta)\Big) + \frac{k_v}{T_e}\big(\dot{v}_L(t-\Delta) - 2\dot{v}_1(t-\Delta) + \dot{v}_2(t-\Delta)\big)$$

$$\dddot{v}_2(t) = \frac{k_d}{T_e}\big(v_1(t-\Delta) - 2v_2(t-\Delta) + v_3(t-\Delta)\big) + \frac{k_d}{T_e}\Big(v_1(t-\Delta) - v_2(t-\Delta) - l - T_{g,des}\dot{v}_2(t-\Delta)\Big) + \frac{k_v}{T_e}\big(\dot{v}_1(t-\Delta) - 2\dot{v}_2(t-\Delta) + \dot{v}_3(t-\Delta)\big)$$

$$\vdots$$

$$\ddot{v}_i(t) = \frac{k_d}{T_e}\big(v_{i-1}(t-\Delta) - 2v_i(t-\Delta) + v_{i+1}(t-\Delta)\big)$$
$$+\frac{k_d}{T_e}\Big(v_{i-1}(t-\Delta) - v_i(t-\Delta) - l - T_{g,des}\dot{v}_i(t-\Delta)\Big)$$
$$+\frac{k_v}{T_e}\big(\dot{v}_{i-1}(t-\Delta) - 2\dot{v}_i(t-\Delta) + \dot{v}_{i+1}(t-\Delta)\big)$$

.
.
.

$$\ddot{v}_N(t) = \frac{k_d}{T_e}\Big(v_{N-1}(t-\Delta) - v_N(t-\Delta) - l - T_{g,des}\dot{v}_N(t-\Delta)\Big) + \frac{k_v}{T_e}\big(\dot{v}_{N-1}(t-\Delta) - \dot{v}_N(t-\Delta)\big) \quad (23)$$

The above representation can be transformed into space gap error coordinates using the following transformations,

$$z_1(t) = p_L(t) - p_1(t) - l - T_{g,des}v_1(t)$$
$$z_2(t) = p_1(t) - p_2(t) - l - T_{g,des}v_2(t)$$

.
.
.

$$z_i(t) = p_{i-1}(t) - p_i(t) - l - T_{g,des}v_i(t)$$

.
.
.

$$z_{N-1}(t) = p_{N-2}(t) - p_{N-1}(t) - l - T_{g,des}v_{N-1}(t)$$
$$z_N(t) = p_{N-1}(t) - p_N(t) - l - T_{g,des}v_N(t) \quad (24)$$

Differentiating both sides of (24) with respect to time thrice,

$$\dddot{z}_1(t) = \ddot{v}_L(t) - \ddot{v}_1(t) - T_{g,des}\dddot{v}_1(t)$$
$$\dddot{z}_2(t) = \ddot{v}_1(t) - \ddot{v}_2(t) - T_{g,des}\dddot{v}_2(t)$$

.
.
.

$$\dddot{z}_i(t) = \ddot{v}_{i-1}(t) - \ddot{v}_i(t) - T_{g,des}\dddot{v}_i(t)$$

.
.
.

$$\dddot{z}_{N-2}(t) = \ddot{v}_{N-2}(t) - \ddot{v}_{N-1}(t) - T_{g,des}\dddot{v}_{N-1}(t)$$
$$\dddot{z}_{N-1}(t) = \ddot{v}_{N-1}(t) - \ddot{v}_N(t) - T_{g,des}\dddot{v}_N(t) \quad (25)$$

Substituting $\ddot{v}_{N-1}(t)$ and $\ddot{v}_N(t)$ according to (22), and $\dddot{v}_N(t)$ from (23) into the expression of $\dddot{z}_{N-1}(t)$ in (25),

$$
\begin{aligned}
\dddot{z}_{N-1}(t) = \dddot{v}_{N-1}(t) - \ddot{v}_N(t) - T_{g,des}\dddot{v}_N(t) \\
&= \frac{k_d}{T_e}\left(p_{N-3}(t-\Delta) - 2p_{N-2}(t-\Delta) + p_{N-1}(t-\Delta)\right) \\
&+ \frac{k_d}{T_e}\left(p_{N-3}(t-\Delta) - p_{N-2}(t-\Delta) - l - T_{g,des}v_{N-2}(t-\Delta)\right) \\
&+ \frac{k_v}{T_e}\left(v_{N-3}(t-\Delta) - 2v_{N-2}(t-\Delta) + v_{N-1}(t-\Delta)\right) \\
&- \frac{k_d}{T_e}\left(p_{N-2}(t-\Delta) - 2p_{N-1}(t-\Delta) + p_N(t-\Delta)\right) \\
&- \frac{k_d}{T_e}\left(p_{N-2}(t-\Delta) - p_{N-1}(t-\Delta) - l - T_{g,des}v_{N-1}(t-\Delta)\right) \\
&- \frac{k_v}{T_e}\left(v_{N-2}(t-\Delta) - 2v_{N-1}(t-\Delta) + v_N(t-\Delta)\right) \\
&- \frac{k_d T_{g,des}}{T_e}\left(v_{N-2}(t-\Delta) - 2v_{N-1}(t-\Delta) + v_N(t-\Delta)\right) \\
&- \frac{k_d T_{g,des}}{T_e}\left(v_{N-2}(t-\Delta) - v_{N-1}(t-\Delta) - T_{g,des}\dot{v}_{N-1}(t-\Delta)\right) \\
&- \frac{k_v T_{g,des}}{T_e}\left(\dot{v}_{N-2}(t-\Delta) - 2\dot{v}_{N-1}(t-\Delta) + \dot{v}_N(t-\Delta)\right)
\end{aligned}
$$

Rearranging the terms on the right side of the above expression,

$$
\begin{aligned}
\dddot{z}_{N-1}(t) = \dddot{v}_{N-1}(t) - \ddot{v}_N(t) - T_{g,des}\dddot{v}_N(t) \\
&= \frac{2k_d}{T_e}\left(p_{N-3}(t-\Delta) - p_{N-2}(t-\Delta) - l - T_{g,des}v_{N-2}(t-\Delta)\right) \\
&- \frac{3k_d}{T_e}\left(p_{N-2}(t-\Delta) - p_{N-1}(t-\Delta) - l - T_{g,des}v_{N-1}(t-\Delta)\right) \\
&+ \frac{k_d}{T_e}\left(p_{N-1}(t-\Delta) - p_N(t-\Delta) - l - T_{g,des}v_N(t-\Delta)\right) \\
&+ \frac{k_v}{T_e}\left(v_{N-3}(t-\Delta) - v_{N-2}(t-\Delta) - T_{g,des}\dot{v}_{N-2}(t-\Delta)\right) \\
&- \frac{2k_v}{T_e}\left(v_{N-2}(t-\Delta) - v_{N-1}(t-\Delta) - T_{g,des}\dot{v}_{N-1}(t-\Delta)\right) \\
&+ \frac{k_v}{T_e}\left(v_{N-1}(t-\Delta) - v_N(t-\Delta) - T_{g,des}\dot{v}_N(t-\Delta)\right) \\
&- \frac{k_d T_{g,des}}{T_e}\left(v_{N-2}(t-\Delta) - v_{N-1}(t-\Delta) - T_{g,des}\dot{v}_{N-1}(t-\Delta)\right)
\end{aligned}
$$

Substituting,

$$
\begin{aligned}
\dddot{z}_{N-1}(t) = \frac{2k_d}{T_e}z_{N-2}(t-\Delta) - \frac{3k_d}{T_e}z_{N-1}(t-\Delta) + \frac{k_d}{T_e}z_N(t-\Delta) + \frac{k_v}{T_e}\dot{z}_{N-2}(t-\Delta) - \frac{2k_v}{T_e}\dot{z}_{N-1}(t-\Delta) \\
+ \frac{k_v}{T_e}\dot{z}_N(t-\Delta) - \frac{k_d T_{g,des}}{T_e}\dot{z}_{N-1}(t-\Delta)
\end{aligned}
$$

Similarly,

$$
\begin{aligned}
\dddot{z}_N(t) = \frac{2k_d}{T_e}z_{N-1}(t-\Delta) - \frac{3k_d}{T_e}z_N(t-\Delta) + \frac{k_d}{T_e}z_{N+1}(t-\Delta) + \frac{k_v}{T_e}\dot{z}_{N-1}(t-\Delta) - \frac{2k_v}{T_e}\dot{z}_N(t-\Delta) \\
+ \frac{k_v}{T_e}\dot{z}_{N+1}(t-\Delta) - \frac{k_d T_{g,des}}{T_e}\dot{z}_N(t-\Delta)
\end{aligned}
$$

Omitting $z_{N+1}(t-\Delta)$ and $\dot{z}_{N+1}(t-\Delta)$ as they require an additional truck following the last or $(N+1)^{th}$ follower truck to exist,

$$\dddot{z}_N(t) = \frac{2k_d}{T_e} z_{N-1}(t-\Delta) - \frac{3k_d}{T_e} z_N(t-\Delta) + \frac{k_v}{T_e}\dot{z}_{N-1}(t-\Delta) - \frac{2k_v}{T_e}\dot{z}_N(t-\Delta) - \frac{k_d T_{g,des}}{T_e}\dot{z}_N(t-\Delta) \tag{26}$$

Without any loss of generality, we assume $z_{N-1}(0) = z_N(0) = 0$. Then, taking the Laplace transformation on both sides of (26),

$$s^3 z_N(s) = \frac{2k_d}{T_e} z_{N-1}(s)e^{-\Delta s} - \frac{3k_d}{T_e} z_N(s)e^{-\Delta s} + \frac{k_v}{T_e} z_{N-1}(s)se^{-\Delta s} - \frac{2k_v}{T_e} z_N(s)se^{-\Delta s} - \frac{k_d T_{g,des}}{T_e} z_N(s)se^{-\Delta s} \tag{27}$$

As mentioned before, a string stable BCM helps propagate and attenuate any perturbations bidirectionally. However, the first and the last vehicles in a BCM are unique because they do not have any immediate leading and following vehicles, respectively. Thus, the string stability analysis of a BCM can be done unidirectionally as the first and the last vehicles have immediate neighbors in only one direction. As mentioned by Eyre et al. [8], the conditions for string stability of a bilateral control can be derived by considering the last two vehicles in a platoon. The space gap error transfer function for the last two trucks can be written from (27) as,

$$G_N(s) = \frac{(2k_d + k_v s)e^{-\Delta s}}{T_e s^3 + (2k_v + k_d T_{g,des})se^{-\Delta s} + 3k_d e^{-\Delta s}} \tag{28}$$

To ensure $\mathcal{L}_\infty$ string stability of the asymmetric LBCM, $|G_N(j\omega)| < 1$, for all $\omega > 0$ [8]. Now,

$$G_N(j\omega) = \frac{(2k_d + jk_v\omega)e^{-j\omega\Delta}}{-jT_e\omega^3 + (2k_v + k_d T_{g,des})j\omega e^{-j\omega\Delta} + 3k_d e^{-j\omega\Delta}}$$
$$= \frac{(2k_d\cos(\omega\Delta) + k_v\omega\sin(\omega\Delta)) + j(k_v\omega\cos(\omega\Delta) - 2k_d\sin(\omega\Delta))}{3k_d\cos(\omega\Delta) + (2k_v + k_d T_{g,des})\omega\sin(\omega\Delta) + j\left((2k_v + k_d T_{g,des})\omega\cos(\omega\Delta) - 3k_d\sin(\omega\Delta) - T_e\omega^3\right)}$$

Then,

$$|G_N(j\omega)|$$
$$= \sqrt{\frac{4k_d^2 + k_v^2\omega^2}{9k_d^2 + (2k_v + k_d T_{g,des})^2\omega^2 + T_e^2\omega^6 + 6k_d T_e\omega^3\sin(\omega\Delta) - (2k_v + k_d T_{g,des})T_e\omega^6\cos(\omega\Delta)}}$$
$$= \sqrt{\frac{X}{X+Y}} \tag{29}$$

where, $X = 4k_d^2 + k_v^2\omega^2$, and $Y = 5k_d^2 + 3k_v^2\omega^2 + k_d^2T_{g,des}^2\omega^2 + 4k_dk_vT_{g,des}\omega^2 + T_e^2\omega^6 + 6k_dT_e\omega^3\sin(\omega\Delta) - (2k_v + k_dT_{g,des})T_e\omega^6\cos(\omega\Delta)$. Clearly, $X > 0$ for all $k_d$, $k_v$, $\omega > 0$. Therefore, $Y > 0$ for all $\omega > 0$ will ensure $|G_N(j\omega)| < 1$. Considering $\sin(\omega\Delta) \le \omega\Delta$, and $\cos(\omega\Delta) \le 1$,

$$Y \ge 5k_d^2 + 3k_v^2\omega^2 + k_d^2T_{g,des}^2\omega^2 + 4k_dk_vT_{g,des}\omega^2 + T_e^2\omega^6 + 6k_dT_e\Delta\omega^4 - (2k_v + k_dT_{g,des})T_e\omega^6$$

Since $5k_d^2 > 0\ \forall k_d > 0$, to ensure $Y > 0$ for all $\omega > 0$ (and so $|G_N(j\omega)| < 1$), we have,

$$-(2k_v + k_dT_{g,des} - T_e)T_e\omega^4 + 6k_dT_e\Delta\omega^2 + (3k_v^2 + k_d^2T_{g,des}^2 + 4k_dk_vT_{g,des}) \ge 0$$

Taking a conservative approach, let us consider,

$$-(2k_v + k_dT_{g,des} - T_e)T_e\omega^4 + 6k_dT_e\Delta\omega^2 + (3k_v^2 + k_d^2T_{g,des}^2 + 4k_dk_vT_{g,des}) = 0 \tag{30}$$

Now, since $\omega^2$ is real, we get from (30),

$$36k_d^2T_e^2\Delta^2 + 4T_e\left(3k_v^2 + k_d^2T_{g,des}^2 + 4k_dk_vT_{g,des}\right)\left(2k_v + k_dT_{g,des} - T_e\right) \geq 0$$

Clearly, $36k_d^2T_e^2\Delta^2$ and $3k_v^2 + k_d^2T_{g,des}^2 + 4k_dk_vT_{g,des} > 0$ for all $k_d$, $k_v$, $T_{g,des}$, $T_e$, $\Delta > 0$. Also, it is only safe to assume that $T_{g,des} > T_e$ so that the trucks do not collide with one another, which implies $2k_v + k_dT_{g,des} - T_e > 0$ $\forall k_d$, $k_v > 0$. Therefore, $36k_d^2T_e^2\Delta^2 + 4T_e\left(3k_v^2 + k_d^2T_{g,des}^2 + 4k_dk_vT_{g,des}\right)\left(2k_v + k_dT_{g,des} - T_e\right) > 0$ always holds, i.e., $Y > 0$, which implies that $|G_N(j\omega)| < 1$ for all $\omega > 0$. As $Y > 0$ and $|G_N(j\omega)| < 1$ for all $\omega > 0$, string stability is always achievable with the asymmetric LBCM for an appropriate choice of $T_{g,des}$ setting given $T_e$ and $\Delta$. However, since it is very challenging to analytically obtain a relation between $T_{g,des}$, $T_e$, and $\Delta$, we use the simulation trials to determine the minimum achievable $T_{g,des}$ given $T_e$ and $\Delta$, which we explain in section 4.4.

## 4 EVALUATION OF THE ASYMMETRIC LBCM WITH A CASE STUDY OF AN AUTOMATED TRUCK PLATOON

We evaluate the performance of the asymmetric LBCM in terms of platoon operational efficiency, i.e., how well the model can maintain a constant desired time gap, and stability by simulating a platoon of six fully automated trucks. In this study, a truck platoon represents a CACC platoon of fully automated trucks. To demonstrate the efficacy of this model, we compare the asymmetric LBCM with the symmetric LBCM.

### 4.1 Simulation Parameters and Evaluation Scenarios

In this study, we simulate a CACC platoon of six automated trucks (one leader and five follower trucks) in MATLAB for a total simulation time of 900 sec. Initially, the follower trucks are assumed to be 5 m away from achieving their desired time gaps. The initial speed of all six trucks is 31.44 m/sec or 70.3 mph. The choice of different speed states (as shown in Table 3) is influenced by typical freeway speed limits in the United States. In the simulation, the input parameters are the number of trucks in the platoon, the total simulation time, the initial speed of the follower trucks, the speed profile of the leader truck, and the powertrain delays and lags. The simulation input parameters are summarized in Table 3.

The traffic states in the simulation are defined by the leader truck's speed profile which is obtained from a calibrated traffic simulation network of the I-26 in Berkeley, Orangeburg, and Dorchester County in South Carolina, developed in [26]. The I-26 roadway network was created in VISSIM traffic simulation software and calibrated based on collected field data to yield simulated volumes and travel times within 10% of the actual volume and travel time data. The details of the network development and calibration can be found in [26]. Note that, in our evaluation scenarios, only the leader truck's speed profile is obtained from a VISSIM simulation, i.e., all the follower trucks operate using the control models.

We use the truck vehicle dynamics described in section 3.3 as an input to the VISSIM I-26 network. We define two reduced speed areas (with speed limits of 55 mph and 45 mph) in the VISSIM I-26 network. Except for the two reduced speed areas, the other portions of the I-26 network have a speed limit of 75 mph. When the trucks enter one of the reduced speed areas, they must immediately reduce their speed and therefore undergo a sharp non-linear deceleration period. As the leader truck's speed profile is selected from one of the trucks of this network, the reduced speed areas define various traffic states based on the leader truck's speed profile. Table 3 summarizes different evaluation scenarios depending on the traffic states, which are defined by the selected truck's speed profile from VISSIM. The following assumptions are made for the evaluation scenarios, (i) all the trucks in the platoon operate on a single lane, (ii) there is no cut-in traffic, and once the platoon is formed, i.e., all the follower trucks achieve the desired time gap with their immediate leading trucks, no trucks attempt to merge with or diverge from the platoon, (iii) the roads do not have any vertical or horizontal curvature, and (iv) the follower trucks receive the location and speed information of their immediate neighboring trucks in real-time without any delay.

Table 3: Summary of simulation scenarios considered in this study

| Input Parameters | Simulation Requirement |
|---|---|
| Number of trucks in the platoon | One leader and five follower trucks |
| Initial speed of the leader truck | 31.44 m/sec (70.3 mph) |
| Initial speed of all the follower trucks | 31.44 m/sec (70.3 mph) |
| Total simulation time | 900 sec |
| Simulation step size | 0.001 sec |
| Powertrain delay | 3 different settings: 0.1 sec, 0.2 sec, and 0.3 sec |
| Powertrain lag | 3 different settings: 0.1 sec, 0.2 sec, and 0.3 sec |
| Evaluation scenarios based on different traffic states defined by the leader truck | Uniform speed or zero acceleration states<br>▪ *State 1*: 31.44 m/sec (70.3mph) from 0 sec to 149 sec, 359 sec to 562 sec, and from 712 sec to 900 sec<br>▪ *State 2*: 19.69 m/sec (44.04 mph) from 158 sec to 240 sec<br>▪ *State 3*: 24.15 m/sec (54.02 mph) from 569 sec to 634 sec<br><br>Speed with non-linear acceleration states<br>▪ *State 1* (from 240 sec to 359 sec): Speed changes from 19.69 m/sec (44.04 mph) to 31.44 m/sec (70.3 mph)<br>▪ *State 2* (from 634 sec to 712 sec): Speed changes from 24.15 m/sec (54.02 mph) to 31.44 m/sec (70.3 mph)<br><br>Speed with non-linear deceleration states<br>▪ *State 1* (from 149 sec to 158 sec): Speed changes from 31.44 m/sec (54.02 mph) to 19.69 m/sec (44.04 mph)<br>▪ *State 2* (from 562 sec to 569 sec): Speed changes from 31.44 m/sec (74.3 mph) to 24.15 m/sec (54.02 mph) |

### 4.2 Control Gain Estimation

To estimate the control gains of the symmetric and asymmetric LBCMs for an automated truck platoon, we focus on minimizing (i) the deviation of the follower trucks' speed from their immediate leading truck's speed, and (ii) the deviation of the follower trucks' actual time gap from the constant desired time gap. For both LBCMs, there are three control gains to estimate: (i) relative distance gain, $k_d$ (for the asymmetric LBCM, we consider $k_{d1} = k_{d2} = k_d$ for simplicity of design, and for the symmetric LBCM, $k_{d1} = k_d$ and $k_{d2}$ is set to zero as the symmetric LBCM does not incorporate the constant desired time gap feature); (ii) relative speed gain, $k_v$, and (iii) feedback gain based on $(v_{des} - v)$, $k_c$.

We use the Genetic Algorithm (GA) for estimating the control gains of the symmetric and asymmetric LBCMs. The GA, developed by Holland, is a metaheuristic optimization method that mimics the process of natural selection and natural genetics [10,12]. The GA can be used for both constrained (i.e., when the lower and upper bounds of the decision variables are specified) and unconstrained (i.e., when the lower and upper bounds of the decision variables are not specified) optimization problems. We apply GA in this study only to estimate the control gains as it is a global optimization algorithm that can globally optimize a multi-objective fitness function while satisfying non-linear inequality constraints, such as the constraints imposed by (21) (presented in section 3.3.1). Besides, GA has been widely used in the literature to calibrate microscopic simulation models [15,18,21,30].

To initialize the GA-based optimization for control gain estimation, we choose a population size of 50 candidate solutions, i.e., in each population, 50 candidate solutions are generated which are evaluated based on the fitness function. As the number of decision variables is less than five for both LBCMs, a population size of 50 candidate solutions is sufficient. We define the two objectives of the multi-objective fitness function ($y$) as follows: (i) $y(1)$: the root mean squared errors (RMSE) of the follower trucks' speed with respect to their immediate leading trucks' speed, and (ii) $y(2)$:

the RMSE of the follower trucks' time gaps with respect to the constant desired time gap ($T_{g,des}$), which can be written as follows,

$$y(1) = (((v_L - v_1)^2 + (v_1 - v_2)^2 + \cdots + (v_{i-1} - v_i)^2 + \cdots + (v_{N-1} - v_N)^2)/N)^{\frac{1}{2}} \quad (31)$$

$$y(2) = \sum_{i=1}^{N} \sqrt{\frac{\left(T_{g,des} - T_{h,i}\right)^2}{N}} \quad (32)$$

where, $N$ is the total number of follower trucks; $v_L$ and $v_i$, are the speeds of the leader truck of the platoon and the $i$-th follower truck, respectively; and $T_{h,i}$ is the actual time gap of the $i$-th follower truck with its immediate leading truck.

We use a separate simplified leader truck speed profile with linear accelerations and decelerations only for the control gain estimation purpose (as shown in Figure 3). As $T_{g,des}$ = 0.8 sec is the most conservative among the different achievable time gap settings (based on the different lumped powertrain delays and lags) used for the evaluation section in this paper, we consider it safe to assume that a set of control gains that satisfy the string stability condition for $T_{g,des}$ = 0.8 sec would also satisfy any other desired time gap setting for platooning that is greater than 0.8 sec.

Table 4 presents a summary of the control gains obtained from the GA optimization that minimizes the fitness function defined in (31) and (32) for the leader truck's speed profile presented in Figure 3.

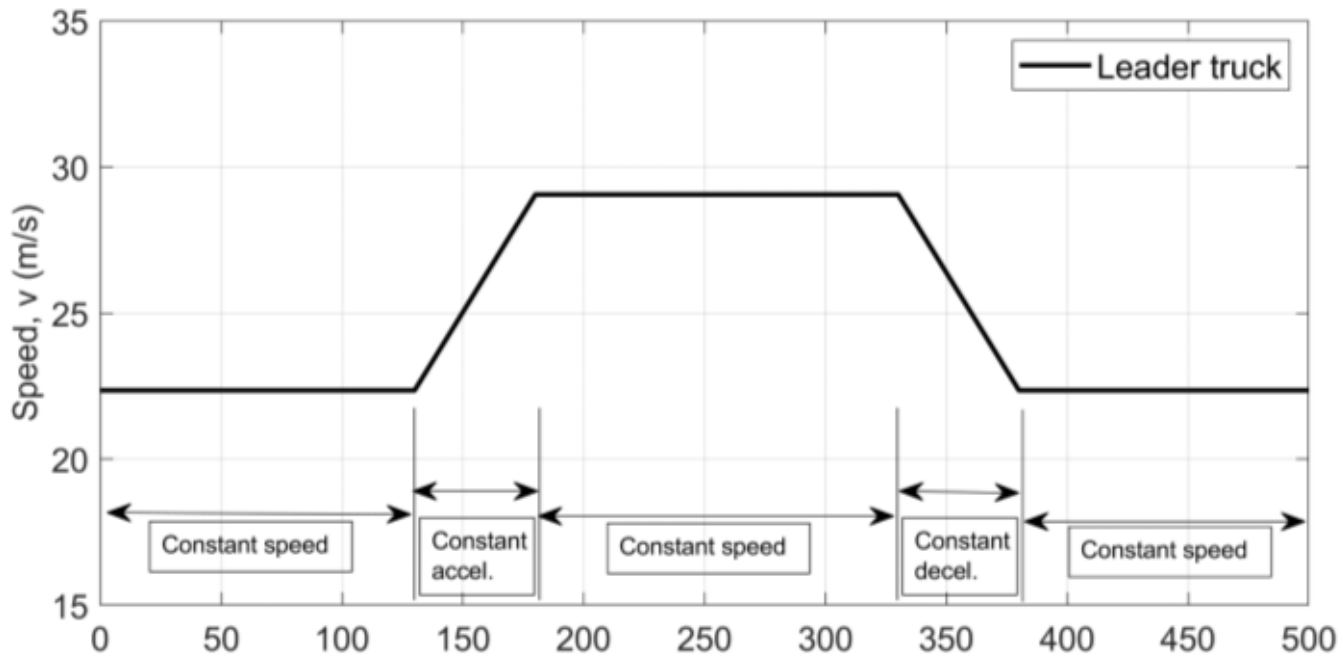

Figure 3: Speed profile of the leader truck for control gain design.

Table 4: Summary of control gains used in this study

| Control Gains | Values |
|---|---|
| **Symmetric LBCM** | |
| Relative distance gain, $k_d$ | 0.8322 sec$^{-2}$ |
| Relative speed gain, $k_v$ | 1.6170 sec$^{-1}$ |
| Feedback gain based on $(v - v_{des})$, $k_c$ | 9.927e-4 sec$^{-1}$ |
| **Asymmetric LBCM** | |
| Relative distance gain, $k_{d1}$ | 1.9589 sec$^{-2}$ |
| Relative gap gain, $k_{d2}$ | 1.9589 sec$^{-2}$ |
| Relative speed gain, $k_v$ | 0.52 sec$^{-1}$ |
| Feedback gain based on $(v_c - v_{des})$, $k_c$ | 0.04 sec$^{-1}$ |

### 4.3 Evaluation Metrics

To numerically evaluate the platoon operational efficiency and the string stability, we introduce two evaluation metrics: (i) the sum of squared time gap error (SSTE), and (ii) the sum of squared speed error (SSSE). SSTE and SSSE at a given timestamp $(t_i)$ are calculated as follows,

$$SSTE(t_i) = \sum_{j=1}^{N} \left(T_{g,j}(t_i) - T_{g,des}\right)^2 \tag{33}$$

$$SSSE(t_i) = \left(v_L(t_i) - v_1(t_i)\right)^2 + \sum_{j=2}^{N} \left(v_{j-1}(t_i) - v_j(t_i)\right)^2 \tag{34}$$

where, $T_{g,j}(t_i)$ is the actual time gap of the $j^{th}$ follower truck at $t_i$ with its immediate leading truck, $v_L(t_i)$ is the leader truck's speed at $t_i$, and $v_j(t_i)$ is the $j^{th}$ follower truck's speed at $t_i$.

We choose to use the sum of the squared errors as it (i) magnifies the deviations of individual follower truck from the desired behavior by squaring, and (ii) combines the errors or deviations of all the follower trucks by summing them up.

SSTE measures how well the follower trucks in the platoon can maintain a tightly coupled platoon formation compared to the constant desired time gap ($T_{g,des}$) when the speed of the leader truck changes. Thus, SSTE is a representation of platoon operational efficiency here. SSTE = 0 at any timestamp, $t_i$ indicates that all the follower trucks in the platoon maintain the desired time gap with their immediate leading trucks exactly without any deviation. A smaller inter-truck gap or time gap is essential to achieve higher platoon operational efficiency. Therefore, for a platoon of trucks, maintaining lower SSTE for a constant desired time gap allows the platoon to achieve higher platoon operational efficiency. A comparison of SSTE among the models can help to determine which model can provide higher platoon operational efficiency for a truck platooning application. For example, if model #1 yields an overall lower SSTE than model #2, then it can be concluded that model #1 provides higher platoon operational efficiency as compared to model #2.

SSSE measures the fluctuation of speed, i.e., SSSE = 0 indicates that all the follower trucks in the platoon follow their immediate leading truck's speed exactly without any errors/deviations. Thus, a lower SSSE of a model compared to another is an indicator of higher string stability. A platoon of trucks can be considered string stable when any non-zero speed error of any truck in the platoon does not get amplified in the upstream, i.e., in the follower trucks [4,24]. Therefore, as SSSE is the sum of squared speed errors of all the follower trucks in the platoon with respect to their immediate leading truck at any given time, a comparison of SSSE profiles among the models can reveal the level of string stability rendered by one model as compared to the other models. For example, if model #1 consistently yields lower SSSE than model #2, then it can be concluded that model #1 renders better string stability compared to model #2.

### 4.4 Evaluation Outcomes

In this subsection, we present the evaluation outcomes of the asymmetric LBCM in terms of platoon operational efficiency, and local and string stability of an automated truck platoon. As explained previously, a CACC platoon of six automated trucks is simulated in a 900-sec simulation scenario with different lumped powertrain delays and lags, i.e., $T_e$ and $\Delta$ = 0.1 sec to 0.3 sec, to investigate the efficacy of the asymmetric LBCM numerically. The platoon of six trucks (one leader truck and five follower trucks) is simulated in MATLAB by solving a system of first-order differential equations. We follow the same procedure as explained by Rahman et al. [25] to form a system of first-order differential equations and use the "ode45" MATLAB solver. The simulation scenarios and the control gains used in the simulation experiment are presented in Tables 3 and 4, respectively.

Figures 4 to 6 present the speed, time gap, SSSE, and SSTE profiles of the trucks in an automated platoon for $(T_e, \Delta)$ = (0.1 sec, 0.1 sec), (0.1 sec, 0.2 sec), and (0.2 sec, 0.1 sec), respectively, where the follower trucks in the platoon use the symmetric LBCM, and the asymmetric LBCM. Figure 4(a) shows the different uniform speed, acceleration, and deceleration states as explained in Table 3 previously. For each $(T_e, \Delta)$, we experimented with different constant desired time gap (i.e., $T_{g,des}$) settings to find the minimum achievable $T_{g,des}$ ensuring an operationally efficient stable platoon operation, i.e., ensuring the minimum deviation in time gap from $T_{g,des}$, while using the asymmetric LBCM. Then, the same minimum achievable $T_{g,des}$'s were used for the symmetric LBCM to compare the performance of the two LBCMs. For example, Figure 7 presents the maximum values of SSTE (which indicates the maximum deviation in time gap from $T_{g,des}$ during the experiment) for $(T_e, \Delta)$ = (0.1 sec, 0.1 sec), (0.1 sec, 0.2 sec), and (0.2 sec, 0.1 sec) with different $T_{g,des}$ settings. We choose the minimum achievable $T_{g,des}$ when the maximum value of SSTE falls under $10^{-2}$ $sec^2$ and remains unchanged as $T_{g,des}$ is increased. For both the LBCMs, $T_{g,des}$ = 0.8 sec is the minimum achievable $T_{g,des}$'s for $(T_e, \Delta)$ = (0.1 sec, 0.1 sec), as shown in Figure 4. The asymmetric LBCM can maintain $T_{g,des}$ = 1.0 sec under $(T_e, \Delta)$ = (0.1 sec, 0.2 sec), and (0.2 sec, 0.1 sec) as shown in Figures 5 and 6. However, the symmetric LBCM cannot maintain any $T_{g,des}$ for any delay and lag over 0.2 sec, i.e., $(T_e + \Delta) > 0.2$ sec, and the stability of the platoon gets disrupted (as observed in Figures 5 and 6). This behavior is reasonable for the symmetric LBCM since it does not account for a constant desired time gap directly into the model. For a low $(T_e + \Delta)$, such as 0.2 sec or lower, the symmetric LBCM is able to perform well enough to maintain a stable platoon formation, but for $(T_e + \Delta) > 0.2$ sec, the symmetric LBCM causes the stability of the platoon to get disrupted completely.

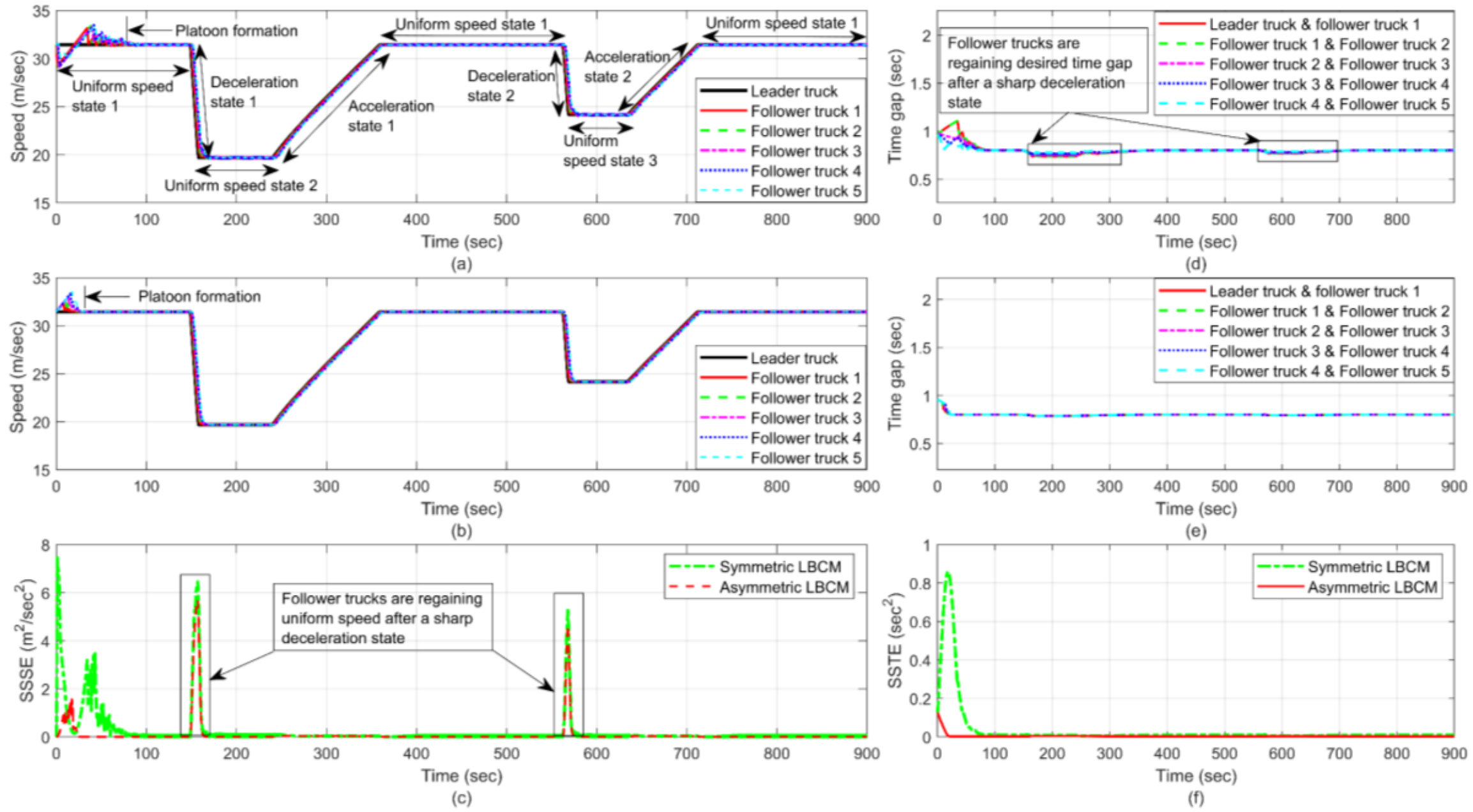

Figure 4: (a) Speed profiles while using the symmetric LBCM, (b) speed profiles while using the asymmetric LBCM, (c) SSSE profile, (d) time gap profiles while using the symmetric LBCM, (e) time gap profiles while using the asymmetric LBCM, and (f) SSTE profile for $T_e$ = 0.1 sec and $\Delta$ = 0.1 sec with $T_{g,des}$ = 0.8 sec.

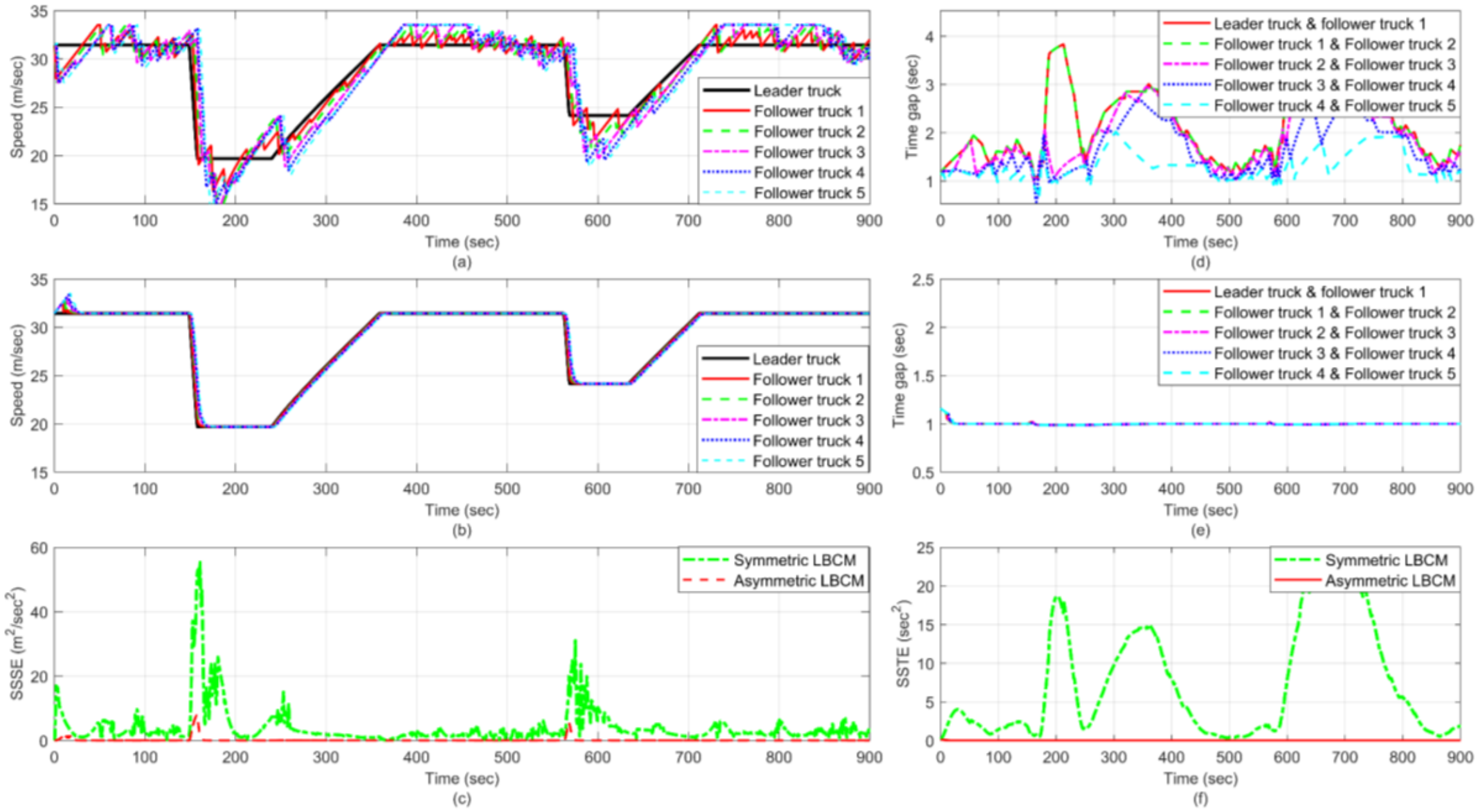

Figure 5: (a) Speed profiles while using the symmetric LBCM, (b) speed profiles while using the asymmetric LBCM, (c) SSSE profile, (d) time gap profiles while using the symmetric LBCM, (e) time gap profiles while using the asymmetric LBCM, and (f) SSTE profile for $T_e$ = 0.1 sec and Δ = 0.2 sec with $T_{g,des}$ = 1.0 sec.

From Figures 4(b), 5(b), and 6(b), we observe the performance of the asymmetric LBCM under non-linear acceleration and deceleration states to evaluate how well this linear model can handle non-linearity imposed by heavy-duty truck vehicle dynamics. The times when the leader truck enters the reduced speed areas and brakes hard to keep its speed within the reduced speed requirement can be considered critical evaluation scenarios. For these short periods of time immediately after the sharp deceleration states, both models cause small fluctuations in speed (i.e., speed fluctuations stay within ±2 m/sec for each follower truck) as observed from the SSSE profiles presented in Figures 4(b), 4(c), but the follower trucks regain uniform speed quickly. As explained before, SSSE is an indicator of string stability. Therefore, both models can ensure string stability for low $(T_e, \Delta)$, such as $(T_e, \Delta)$ = (0.1 sec, 0.1 sec). However, the symmetric LBCM causes the follower trucks to deviate from the constant desired time gap for a longer period of time (as seen in Figures 4(d)).

Figures 5 and 6 reveal the efficacy of the asymmetric LBCM for automated truck platooning compared to the symmetric LBCM. As $(T_e + \Delta) > 0.2$ sec, the symmetric LBCM fails to ensure a stable platoon operation, i.e., the follower trucks fail to follow the leader truck's speed profile (as seen from Figures 5(a) and 6(a)) as well as maintain $T_{g,des}$ (as seen in Figures 5(d) and 6(d)). However, the asymmetric LBCM remains consistent in terms of closely following the leader truck's speed profile as well as maintaining $T_{g,des}$ as seen in Figures 5(b) and 6(b), and 5(e) and 6(e), respectively. Since the asymmetric LBCM can provide an overall lower SSTE compared to the symmetric LBCM (see Figures 5(f) and 6(f)), an automated truck platoon using the asymmetric LBCM can be operationally more efficient.

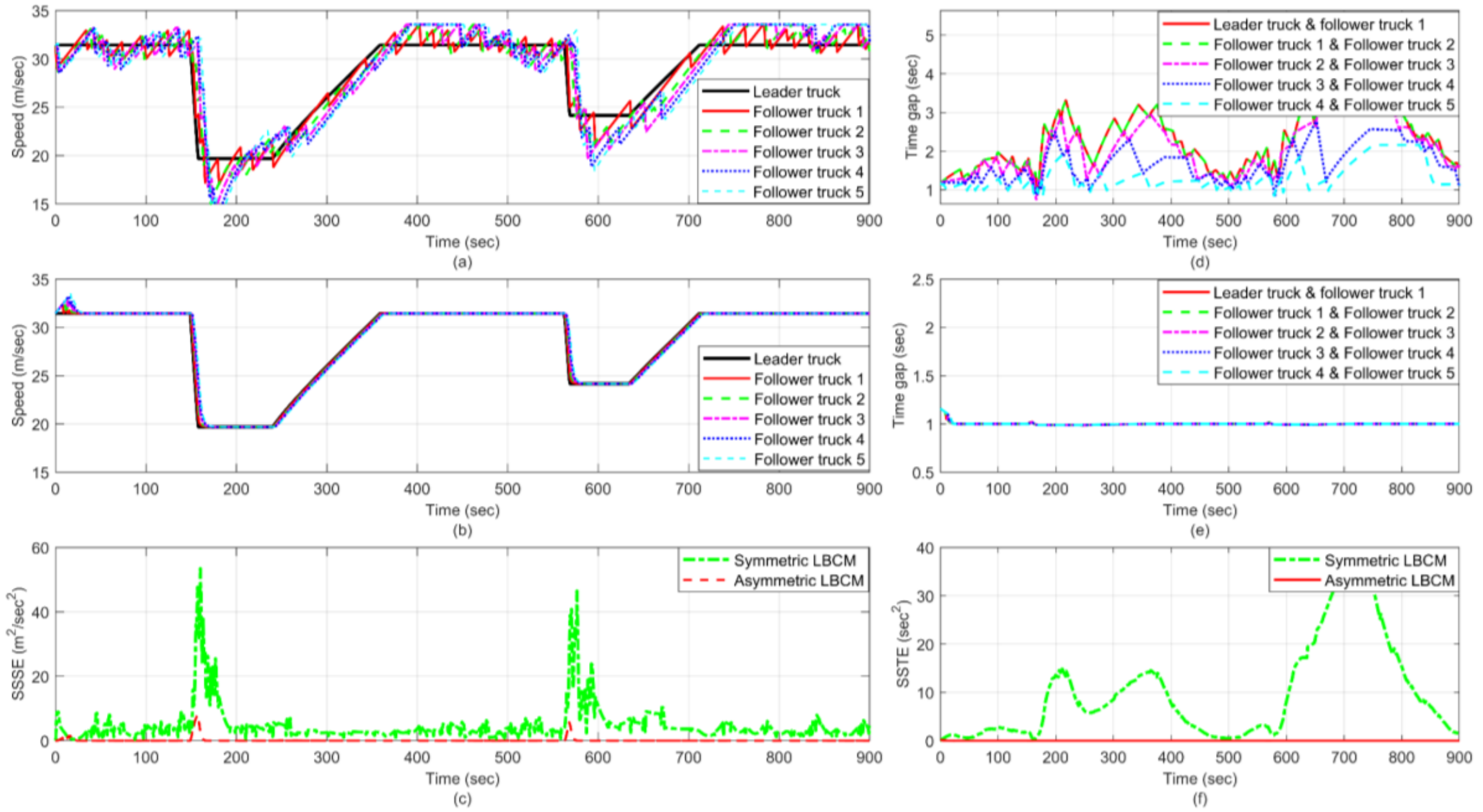

Figure 6: (a) Speed profiles while using the symmetric LBCM, (b) speed profiles while using the asymmetric LBCM, (c) SSSE profile, (d) time gap profiles while using the symmetric LBCM, (e) time gap profiles while using the asymmetric LBCM, and (f) SSTE profile for $T_e$ = 0.2 sec and $\Delta$ = 0.1 sec with $T_{g,des}$ = 1.0 sec.

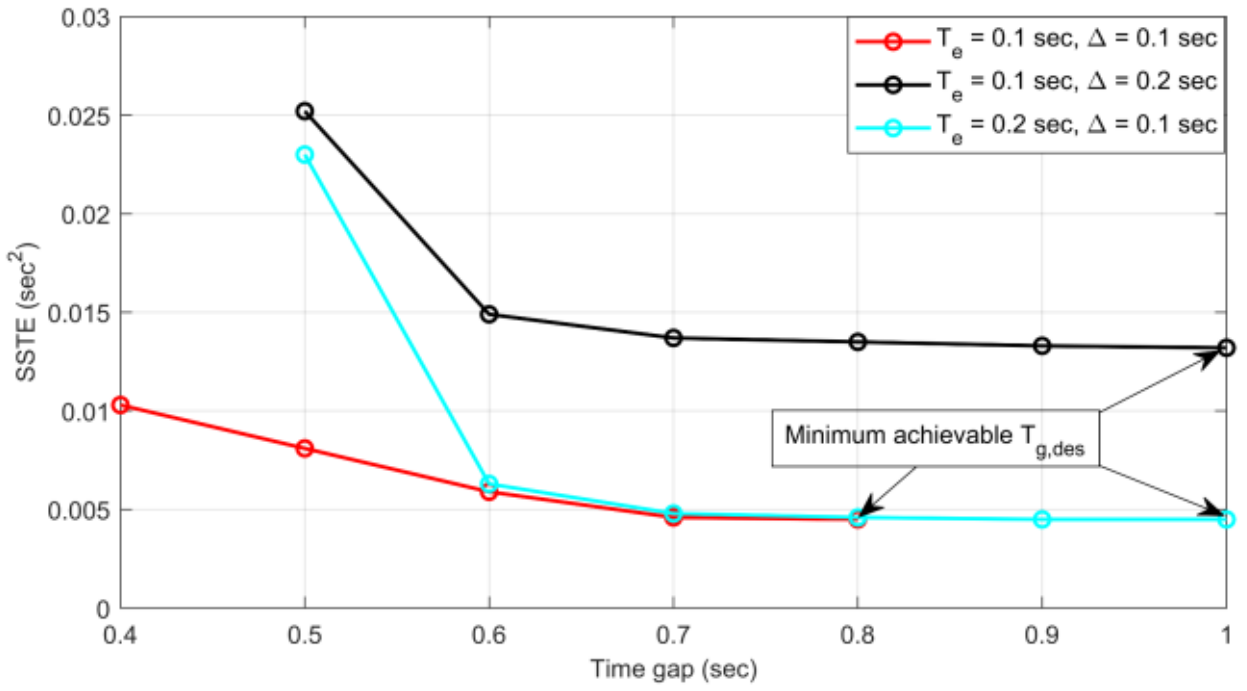

Figure 7: Maximum values of SSTE against different $T_{g,des}$'s for $T_e$ = 0.1 sec and 0.2 sec.

To test the performance of the asymmetric LBCM for even higher $(T_e + \Delta)$'s, we increase both $T_e$ and $\Delta$ up to 0.3 sec with a 0.1-sec increment. The corresponding speed and time gap profiles are presented in Figures 8 to 11. For $(T_e, \Delta)$ = (0.2 sec, 0.2 sec), (0.2 sec, 0.3 sec), (0.3 sec, 0.2 sec), and (0.3 sec, 0.3 sec), the minimum achievable $T_{g,des}$ are 1.5 sec, 1.9 sec, 2.1 sec, and 2.5 sec. As observed in Figures 8 to 11, all the follower trucks are still able to maintain stable platoon formation for increased $(T_e, \Delta)$'s. The follower trucks slightly deviate from $T_{g,des}$ for two short periods of time when the leader truck brakes hard to follow the reduced speed limit (as seen in Figures 8(b) to 11(b)). These deviations from $T_{g,des}$ increase slightly as we increase $T_e$, because, as the leader truck brakes hard and the follower trucks attempt to follow that, the increased $T_e$ causes more delay in braking that accumulates over time during the deceleration states. However, as soon as the leader truck starts operating at reduced uniform speeds during the uniform speed states 2 and 3, the follower trucks are able to regain $T_{g,des}$ quickly. This proves the operational efficiency of the asymmetric LBCM for a stable automated truck platooning operation in terms of closely maintaining a predefined constant desired time gap regardless of the high delays and lags associated with the powertrain system, and limited acceleration and deceleration capabilities experienced by heavy-duty trucks.

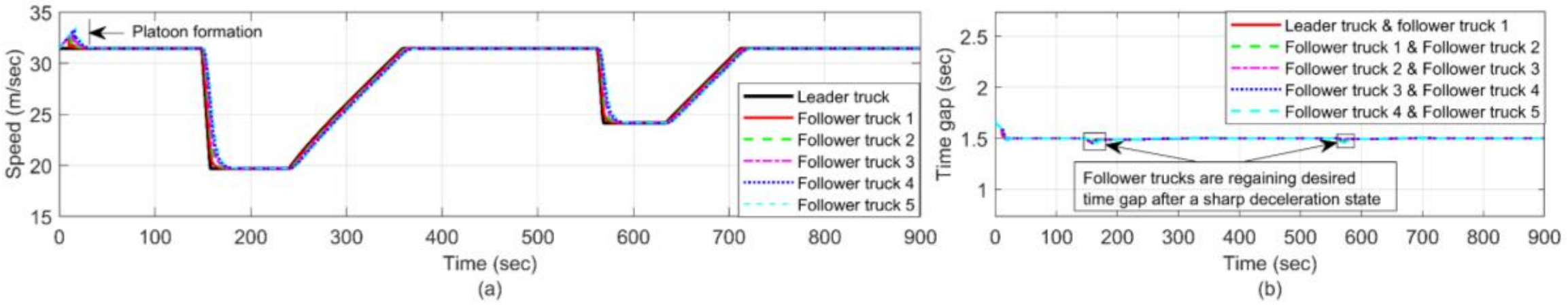

Figure 8: (a) Speed and (b) time gap profiles while using the asymmetric LBCM for $T_e$ = 0.2 sec and Δ = 0.2 sec with $T_{g,des}$ = 1.5 sec.

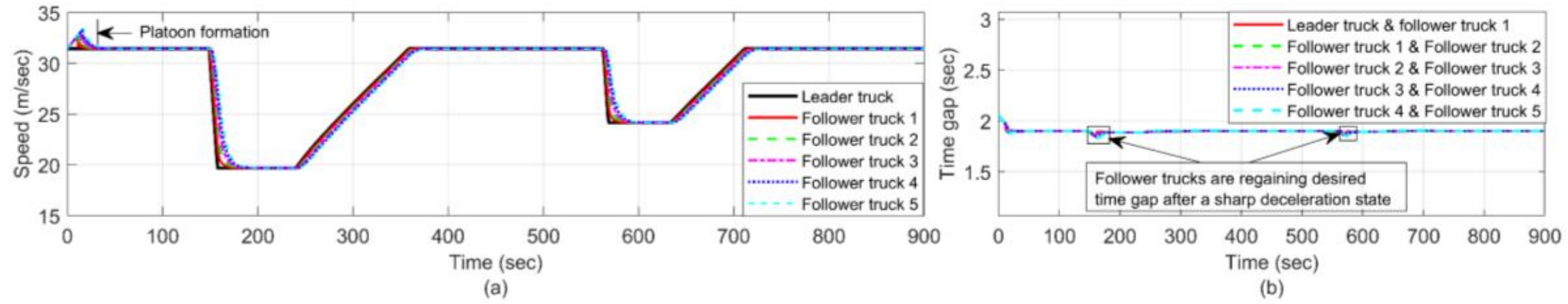

Figure 9: (a) Speed and (b) time gap profiles while using the asymmetric LBCM for $T_e$ = 0.2 sec and Δ = 0.3 sec with $T_{g,des}$ = 1.9 sec.

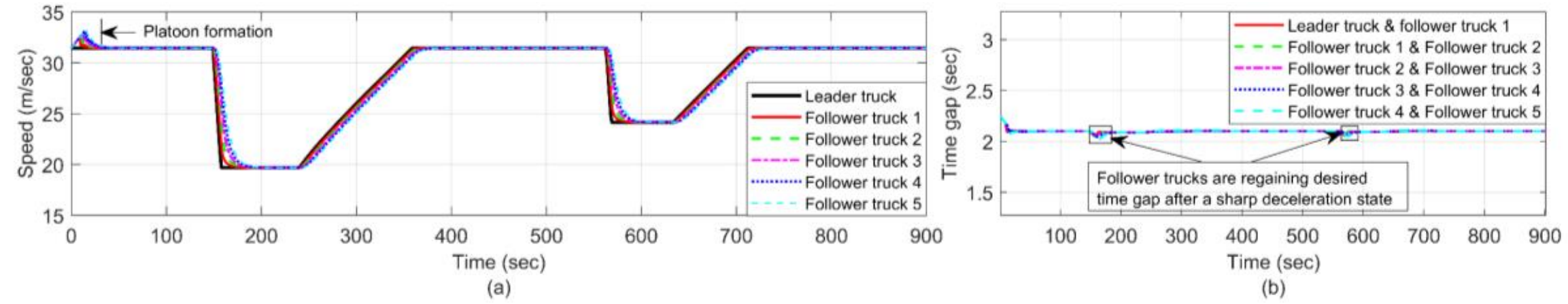

Figure 10: (a) Speed and (b) time gap profiles while using the asymmetric LBCM for $T_e$ = 0.3 sec and Δ = 0.2 sec with $T_{g,des}$ = 2.1 sec.

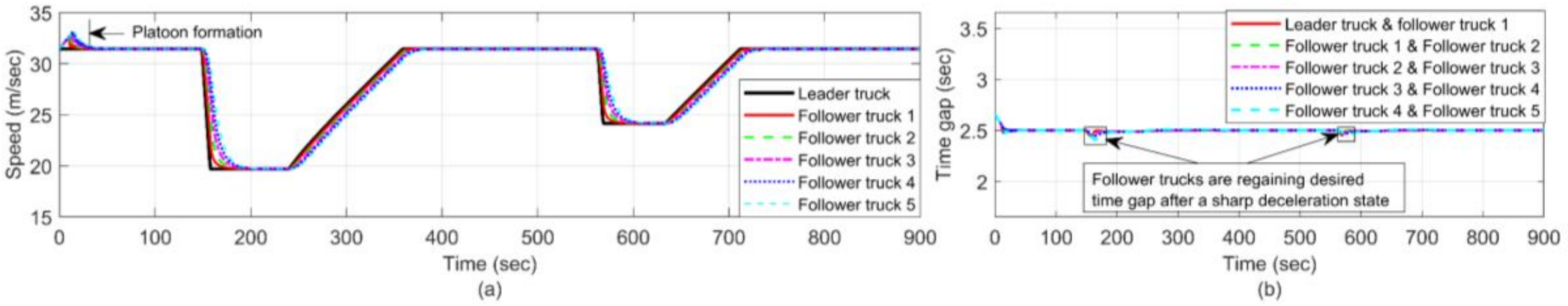

Figure 11: (a) Speed and (b) time gap profiles while using the asymmetric LBCM for $T_e$ = 0.3 sec and Δ = 0.3 sec with $T_{g,des}$ = 2.5 sec.

We further validate the local stability of the asymmetric LBCM by introducing perturbations in the leader truck's speed profile. We consider the smallest and highest combinations of $T_e$ and Δ's with the corresponding $T_{g,des}$'s used in this paper, i.e., $(T_e, \Delta)$ = (0.1 sec, 0.1 sec), and (0.3 sec, 0.3 sec), respectively, for this local stability experiment. As mentioned before, a platoon of trucks using a BCM can be considered locally stable if any perturbation imposed by a truck in the platoon does not cause an increase in speed fluctuations over time for its upstream (i.e., follower) trucks and downstream (i.e., leading) trucks. As shown in Figure 12, the perturbations imposed by the leader truck's speed do not cause an increase in the speed fluctuations of the follower trucks over time, i.e., the perturbations do not make the platoon unstable. Therefore, the asymmetric LBCM ensures local stability as well as string stability (as explained earlier) during platoon operation.

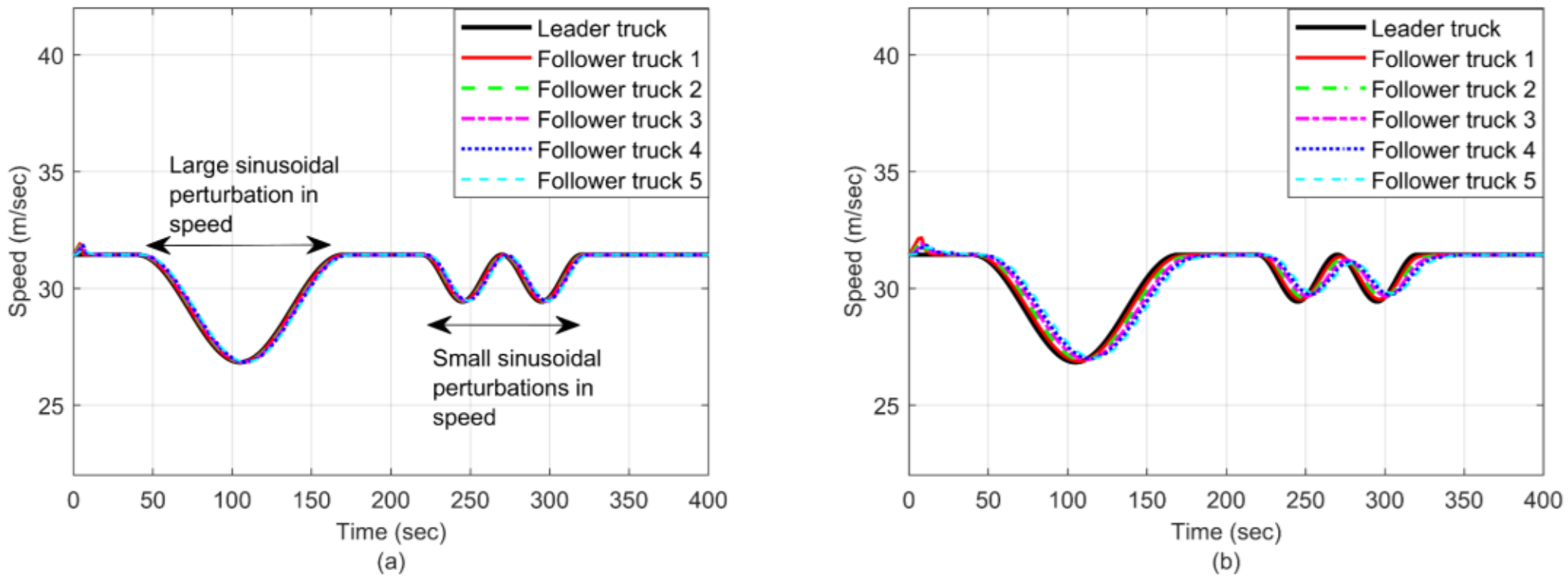

Figure 12: Local stability test using the asymmetric LBCM for (a) $T_e$ = 0.1 sec and Δ = 0.1 sec with $T_{g,des}$ = 0.8 sec, and (b) $T_e$ = 0.3 sec and Δ = 0.3 sec with $T_{g,des}$ = 2.5 sec.

## 5 CONCLUSION

In this paper, we develop an asymmetric LBCM that enables a platoon of fully automated trucks to closely maintain a given constant desired time gap. First, we analyze the stability (local and string stability) of the asymmetric LBCM theoretically. The local stability of the model is analyzed theoretically using the condition for asymptotic stability of an LTI system in the sense of Lyapunov, which provides us a condition involving the desired time gap, lumped powertrain delay and lag, and other control gains. We use this condition to optimize the control gains in order to minimize the follower trucks' deviations from their corresponding desired time gaps and speeds. For analyzing the string stability of the model, we use the space gap error attenuation condition to determine the condition of $\mathcal{L}_\infty$ string stability is achievable under various desired time gap settings. The desired time gap settings to ensure string stability under various powertrain delays and lags are then determined using simulated experiments. Fo evaluation, we numerically investigate the efficacy of the asymmetric LBCM compared to the symmetric LBCM in terms of platoon operational efficiency and stability by simulating a CACC platoon of six automated trucks. To mimic the real-world freeway operation of trucks, different acceleration and deceleration states, such as uniform speed states (i.e., zero acceleration states), non-linear acceleration and deceleration states, and powertrain delays and lags are considered to evaluate the operational performance of an automated truck platoon that uses the asymmetric LBCM.

Our analyses reveal that the asymmetric LBCM can capture the non-linear acceleration and deceleration states as well as handle a wide range of powertrain delays and lags. Each truck in the platoon that uses the asymmetric LBCM can closely follow the speed of the leader truck as well as maintain the constant desired time gap under all simulated scenarios, whereas the trucks in the platoon that uses the symmetric LBCM fail to do that for any combined powertrain delays and lags over 0.2 sec. In addition, the SSTE and the SSSE are estimated, for both models, to numerically compare them for the level of platoon operational efficiency and string stability, respectively. Analyses reveal that the asymmetric LBCM has the minimum SSTE and SSSE compared to the symmetric LBCM under all simulated scenarios. Consequently, it can be concluded that an automated truck platoon that uses the asymmetric LBCM provides better string stability compared to the symmetric LBCM, especially for trucks experiencing high powertrain delays and lags. We experiment with different combinations of powertrain delays and lags for the asymmetric LBCM and determine the minimum achievable desired time gaps for different combined powertrain delays and lags up to 0.6 sec. Further, we evaluate the local stability of an automated truck platoon using the asymmetric LBCM by applying small and large perturbations in the leader truck's speed profile, which shows that the follower trucks can maintain local stability even under the effect of large powertrain delays and lags since the control gains are designed based on the theoretically derived local stability condition.

A limitation of this study is that it focuses on an automated truck platoon formation on a lane without considering trucks moving in and out of the platoon. Our future study will focus on integrating the asymmetric LBCM with trucks moving in and out of the platoon.

## 6 HISTORY DATES